\documentclass[journal]{IEEEtran}
\usepackage{mathpazo}
\usepackage{times}

\usepackage[utf8]{inputenc} 
\usepackage[T1]{fontenc}
\usepackage{url}
\usepackage{ifthen}
\usepackage{cite}
\usepackage{graphicx}
\usepackage{amssymb}
\usepackage[cmex10]{amsmath} 
\usepackage{amsthm}
\usepackage{color}
\usepackage[ruled,vlined,linesnumbered]{algorithm2e}
\usepackage{comment}
\usepackage{caption}

\usepackage{tikz}
\usepackage{pgfplots}

\renewcommand{\le}{\leqslant}

\usepackage{amsthm}

\newtheorem{example}{Example}

\newcommand{\cA}{{\cal A}}

\newcommand{\cL}{{\cal L}}

\newcommand{\cX}{{\cal X}}
\newcommand{\cY}{{\cal Y}} 
\newcommand{\cZ}{{\cal Z}}

\newcommand{\bfc}{{\boldsymbol c}}

\newcommand{\bfu}{{\boldsymbol u}}
\newcommand{\bfv}{{\boldsymbol v}}

\newcommand{\Arikan}{Ar{\i}\-kan}

\begin{document}
\title{Polar Coded Modulation via Hybrid Bit Labeling}
\author{
\IEEEauthorblockN{
    Hanwen Yao\IEEEauthorrefmark{1},
	Jinfeng Du\IEEEauthorrefmark{2}, and
	Alexander Vardy\IEEEauthorrefmark{1}
	}\\[0.27ex]
\IEEEauthorblockA{\IEEEauthorrefmark{1}\small 
Department of Electrical \& Computer Engineering, University of California San Diego, 
La Jolla, CA 92093, USA
} \\[1mm]
\IEEEauthorblockA{\IEEEauthorrefmark{2}\small 
Nokia Bell Labs, Murray Hill, NJ 07974, USA
} \\[0.27ex]
\IEEEauthorblockA{\small Emails: hwyao@ucsd.edu, 
jinfeng.du@nokia-bell-labs.com, avardy@ucsd.edu}\\[-2mm]
\vspace{-5mm}
}

\maketitle

\begin{abstract}
Bit-interleaved coded modulation (BICM) and 
multilevel coded modulation (MLC) are commonly used 
to combine polar codes with high order modulation. 
While BICM benefits from simple design
and the separation of coding and modulation, 
MLC 
shows better performance under successive-cancellation decoding. 
In this paper 
we propose a hybrid polar coded modulation scheme 
that lies between BICM and MLC, 
wherein a fraction of bits are assigned to set-partition
(SP) labeling and the remaining bits are assigned for Gray labeling. 
The SP labeled bits undergo sequential demodulation, 
using iterative demodulation and polar decoding similar 
to MLC, 
whereas the Gray labeled bits are first 
demodulated in parallel and then sent for decoding 
similar to BICM. 
Either polar codes or other channel codes 
(such as LDPC codes) can be used
for the Gray labeled bits. 
For length 2048 rate 1/2 polar 
code on 256-QAM, the performance gap between 
BICM (Gray labeling only)
and MLC (SP labeling only)
can be almost fully closed by the hybrid scheme.
Notably, the hybrid scheme has a significant latency advantage over MLC.
These performance gains 
make the proposed scheme attractive for future
communication systems such as 6G.
\end{abstract}


\section{Introduction}\label{sec:intro}
\noindent 
Polar codes, 
pioneered by Erdal \Arikan \cite{arikan2009channel}, 
are the first family of 
error-correcting codes that provably achieve 
capacity for a wide range of channels, 
with low encoding and decoding complexity.
At short block lengths, 
concatenated with cyclic redundancy check (CRC) outer codes, 
polar codes under successive cancellation list decoding 
\cite{tal2015list}
show competitive, and in some cases, 
better performance as compared with turbo and LDPC codes 
\cite{cocskun2019efficient}. 
Thus polar codes were 
adopted as the error correcting code for control channels
in the fifth generation (5G) 
wireless communications standard \cite{3gpp-5g}.

To achieve higher spectrum efficiency required by the next 
generation wireless networks, it is essential to combine 
polar coding with high order modulation. 
Two commonly used schemes that combine 
polar code with channel modulation are 
bit-interleaved coded modulation (BICM) 
\cite{caire1998bit,i2008bit},
and multilevel coded modulation (MLC) 
\cite{imai1977new,wachsmann1999multilevel}.

In bit-interleaved polar coded modulation (BI-PCM)
\cite{seidl2013polar}, 
polar coding and modulation are connected 
by an interleaver, 
and Gray labeling is commonly used for mapping between 
the coded bits and the constellation symbols.
At the receiver, 
on a constellation with $2^m$ symbols, 
the demodulator computes  the soft information for 
all $m$ bits of each received symbols in parallel, 
which are then de-interleaved 
and passed to the polar decoder.
In BI-PCM, the $2^m$-ary channel is effectively 
decomposed into $m$ binary sub-channels, 
and decoded 
regardless of their dependency.  
Benefits from its easiness of code design 
and the separation of coding and modulation, 
BI-PCM has been adopted for polar code in 
the 5G wireless communication standard \cite{bioglio2020design}.
For constellation whose order $m$ is not a power of 2, 
an additional polarization matrix can be used 
to connect polar codes with channel modulation
\cite{mahdavifar2015polar}. 
However, the major drawback of
BICM is that it is unable to achieve the constellation-constrained
capacity over additive white Gaussian noise (AWGN)
channels \cite{caire1998bit}, 
due to loss of mutual information between the 
decomposed sub-channels.

It is known that MLC together with multi-stage decoding (MSD)  
can achieve the constellation constrained capacity over 
AWGN channels, 
provided that the code rate of each level 
is properly designed \cite{wachsmann1999multilevel}. 
It turns out that MSD is very similar 
to successive cancellation (SC) decoding 
of polar code on the conceptual level. 
Seidl \textit{et al.} \cite{seidl2013polar} 
first discuss the multilevel polar coded modulation (ML-PCM). 
They introduce a \textit{channel parition} framework 
that unifies both the bit channel formation 
arise with SC decoding, 
and the channel decomposition in MSD. 
This framework makes it possible to assign the code rate, 
and design the polar code
at each level of ML-PCM in a consistent way.
In ML-PCM on a constellation with $2^m$ symbols, 
the $2^m$-ary channel is decomposed 
into $m$ binary sub-channels preserving their dependency.
At the receiver, 
a multi-stage demodulator sequentially 
computes the soft information of those $m$ sub-channels 
for each symbol.
At each level, the computation is based on both 
the channel output and the hard values of 
all the previous sub-channels, where the hard values 
are obtained from the polar decoder. 
In this way, 
the mutual information between the 
sub-channels is preserved, 
and ML-PCM is expected to have a better performance 
compared with BI-PCM over AWGN channels. 

However, there are multiple rounds of information exchanges between 
the demodulator and the decoder during its multi-stage decoding process in ML-PCM. 
At each level, 
the demodulator needs to send the evaluated soft information 
for a certain sub-channel over all received symbols 
to the decoder, 
and wait for the hard values from the decoder 
to proceed to the next stage.
This frequent communication 
could introduce considerable latency for the 
ML-PCM receiver. 
To mitigate this latency issue, 
we introduce a hyrbid polar coded modulation design 
that lies between ML-PCM and BI-PCM, 
that is able to reduce the amount of communication
between the demodulator and the decoder, 
while still maintaining a considerable performance gain 
over BI-PCM.
\subsection{Our Contribution}
\noindent
In this paper we propose a new polar coded modulation scheme, referred as hybrid polar coded modulation (Hybrid-PCM) hereafter,
that can be viewed as a comprehensive framework
having both BI-PCM and ML-PCM as its special cases.
In our Hybrid-PCM scheme, 
the $2^m$-ary channel on a 
constellation with $2^m$ symbols is decomposed into 
$m$ binary sub-channels following 
a new channel transform that we refer as
{\it hybrid binary partition}. 
This channel transform has an integer-valued splitting parameter $s$ 
that lies between 0 and $m$. 
At the receiver side, 
the demodulator computes the soft information 
for the the first $s$ sub-channels 
sequentially, based on both the channel output and the hard value of their previous sub-channels. 
Then with the hard information of the first $s$ levels, 
the demodulator estimates 
the rest of the $(m-s)$ sub-channels parallelly
regardless of their dependency.
Vaguely speaking, 
out of those $m$ levels for every received symbol, 
the first $s$ levels are sequentially decoded 
similar to ML-PCM, 
and the last $(m-s)$ levels are parallelly 
decoded similar to BI-PCM.
In this way, only the mutual information of the first $s$ 
sub-channels is preserved during the decoding process, 
and we are free to choose this splitting parameter $s$ 
between $0$ and $m$.
We also propose a hybrid labeling rule to fit our scheme.
This labeling rule lies between Gray labeling, 
commonly used in BICM, 
and set-partitioning (SP) labeling, 
commonly used in MLC, 
and it's governed by the same splitting parameter $s$.

If we choose $s$ to be equal to 0, then our hybrid scheme 
becomes BI-PCM. And if we choose $s$ to be $m$, 
our hybrid scheme becomes ML-PCM. 
For $s$ lying between 0 and $m$, 
our hybrid scheme can reduce the amount of back-and-forth 
communication required in ML-PCM, 
while as we will show in Section V, 
still holding a considerable performance gain over BI-PCM.

%
\section{Preliminaries}
\label{sec:prelim}
\noindent
In this section, we describe our system model, 
and give a brief 
review on the concepts of polar codes 
and polar coded modulation. 
This prepare us for the development of our 
proposed hybrid polar coded modulation scheme. 

Here are some notation conventions 
that we follow throughout this paper. 
We use bold letters like $\bfu$ to denote vectors, 
and non-bold letters like $u_i$ to denote 
symbols within that vector.
For $\bfu = (u_1,u_2,\cdots,u_{n})$, 
we denote its subvector consists of 
symbols with indices from $a$ to $b$ as $\bfu_a^b = (u_a,u_{a+1},\cdots,u_b)$. 
And we use $(\bfu,\bfv)$ to 
denote the concatenation of vector $\bfu$ 
and vector $\bfv$. 

\subsection{System Model}
\noindent
In this paper, we consider memoryless AWGN channels 
with  
quadrature amplitude modulation (QAM) and pulse-amplitude modulation (PAM).
Since any $2^{2m}$-QAM 
constellation can be constructed from two independent 
$2^m$-PAM constellations for the I-channel and Q-channel, 
henceforth we regard every QAM symbol as two independent PAM 
symbols.

For a PAM constellation with $2^m$ symbols, 
its signal points are given by 
$\cX = \{\pm1,\pm3,\cdots,\pm(2^m-1)\}.$
Each symbol in the constellation is labeled by a binary $m$-tuple, 
and we say that symbols in this constellation have $m$ {\it bit levels}. 
The input-output relation of the AWGN channel is given by 
$y = x + z$, 
with $x\in\cX$ for each channel use, 
and $z$ being a zero mean Gaussian noise 
with standard deviation $\sigma_z$. 
The quality of the channel is measured 
by the signal to noise ratio (SNR):
$$SNR = E[x^2]/\sigma_z^2.$$

%
%
\subsection{Polar Codes}
\label{subsec:polar_code}
\noindent
Assuming $n = 2^\ell$, an $(n,k)$ polar code is a binary linear block 
code generated by $k$ rows of the 
polar transformation matrix 
$G_n = K_2^{\otimes \ell}$, where 
\vspace{-3mm}
$$
K_2 = \begin{bmatrix}
1 & 0 \\ 1 & 1
\end{bmatrix},
$$
and $K_2^{\otimes \ell}$ is the $\ell$-th Kronecker power of $K_2$.
The encoding scheme is given
by $\bfc = \bfu G_n$, where $\bfu$ is a
length-$n$ binary input vector carrying $k$ data bits, and
$c$ is the codeword for transmission.
The positions of the data bits in $\bfu$ are specified by 
an information index set $\cA\subseteq\{1,2,\cdots,n\}$ of size $k$, 
with the rest of the $n-k$ bits in $\bfu$ frozen 
to certain fixed values, usually zeros.
The construction for polar codes usually refers to the 
selection of the information index set $\cA$.

For decoding of polar code,
in this paper we consider the conventional SC decoder, 
which is proven to be capacity achieving \cite{arikan2009channel}. 
For details of SC decoding for polar code, 
we refer the readers to Ar{\i}kan's 
seminal paper \cite{arikan2009channel}.

\subsection{Bit-Interleaved Polar Coded Modulation (BI-PCM)}
\label{subsec:BIPCM}
\noindent
Let $|\cX| = 2^m$, and let $N$ denotes the number of channel uses. 
In BI-PCM, 
the binary codeword generated by the polar encoder 
is permuted by an interleaver. 
Then, each block of $m$ bits is mapped into 
a constellation symbol in $\cX$ for channel transmission.
At the receiver's side of BI-PCM, for each received symbol, 
the demodulator ignores the relation between bit levels, 
and computes the soft information for all bit levels 
solely based on the channel observation. 

Let $W:\cX\rightarrow\cY$ 
be a $2^m$-ary channel with input symbol set $\cX$ with $|\cX| = 2^m$, 
and output alphabet $\cY$. 
In a BI-PCM scheme over this channel, 
$W$ is decomposed into $m$ binary sub-channels that 
are viewed as independent channels by the receiver.
This channel transform is called 
{\it parallel binary partition} (PBP) 
in \cite{seidl2013polar}. Here we denote it as
$$
\varphi:\; W\rightarrow\{
B_{\varphi}^{(1)}, 
B_{\varphi}^{(2)}, 
\cdots, 
B_{\varphi}^{(m)} 
\},
$$
where 
$B_{\varphi}^{(j)}:\{0,1\}\rightarrow\cY$ 
denotes the binary sub-channel for the $j$-th bit level 
for $j = 1,2,\cdots,m$. 
In PBP, each sub-channel $B_{\varphi}^{(j)}$ only has 
the knowledge of the channel output $y\in\cY$. 
And Gray labeling is commonly used to generate 
sub-channels that are as independent as possible 
\cite{stierstorfer2007gray}.
Let the bit-to-symbol labeling rule given by 
$\cL:\{0,1\}^m\rightarrow \cX$, 
then $B_{\varphi}^{(j)}$ has the transition probability
$$
B_{\varphi}^{(j)}(y|b) = \frac{1}{2^{m-1}}\sum_{b_1^m\in\{0,1\}^m:\;b_j=b} 
W(y|\cL(b_1^m)), 
$$
for $j = 1,2,\cdots,m$. 

%

After demodulation, 
as shown in Figure \ref{fig:MLPCM_receiver} (left), 
the soft information of all $mN$ bits 
is de-interleaved, and fed to the decoder. 
Note that to use a single polar decoder for BI-PCM, 
the order $m$ of the constellation has to be a power of 2.

\subsection{Compound Polar Code}
\noindent
In BI-PCM, to handle constellation whose order $m$ 
is not necessarily a power of 2, 
{\it compound polar code} is proposed in \cite{mahdavifar2015polar} 
that uses an additional $m\times m$ polarization matrix to 
connect polar code with channel modulation. 
This structure is also mentioned in 
\cite[Sec.V.D]{seidl2013polar}, 
and later used in \cite{chen2013efficient} on 64-QAM.

In BI-PCM with compound polar code, 
the $2^m$-ary channel $W$ is also decomposed into 
$m$ binary sub-channels following PBP, 
but the decomposed channels are not decoded in parallel.
In compound polar code,
an $m\times m$ polarization matrix is used to further 
polarize those $m$ decomposed sub-channels.
And on the receiver's side, 
the polarized channels are decoded sequentially 
based on the hard information of their previous channels.
For the details of compound polar code, 
we refer the readers to \cite{mahdavifar2015polar}.

With this additional polarization matrix, 
compound polar code shows better performance compared 
with plain BI-PCM under SC decoding \cite{mahdavifar2015polar}.
It inherits the benefit that demodulation and 
decoding are separated just like plain BI-PCM, 
but it also introduces extra decoding latency due to that 
additional polarization matrix.
%
%
Since in this paper, 
we focus on reducing the 
iterative communication between the demodulator 
and the decoder in ML-PCM, 
we also include compound polar code 
in our simulation comparison in Section IV.

For our simulation in Section IV, 
we use $K_3$ as the additional polarization matrix for 8-PAM 
the same as in \cite[Sec.VII.A]{mahdavifar2015polar}, 
and use $K_4$ as the additional polarization matrix 
for 16-PAM the same as in \cite[Sec.V.D]{seidl2013polar}:
$$
K_3 = 
\begin{bmatrix}
1 & 0 & 0 \\
1 & 1 & 0 \\
0 & 1 & 1
\end{bmatrix}, 
\qquad
K_4 = 
\begin{bmatrix}
1 & 0 & 0 & 0\\
1 & 1 & 0 & 0\\
0 & 1 & 1 & 0\\
0 & 0 & 1 & 1\\
\end{bmatrix}
$$
It has been shown in \cite{bocherer2017efficient} that 
for the labeling in BI-PCM with compound polar code, 
the least significant bit Binary Reflected Gray Code 
(LSB-BRGC) shows a better performance compared with 
the Binary Reflected Gray Code (BRGC) 
under SC decoding. 
Thus we also adopt LSB-BRGC for the bit labeling 
for BI-PCM with compound polar code in our simulation.
%
%
\subsection{Multilevel Polar Coded Modulation (ML-PCM)}
\label{subsec:MLPCM}
\noindent
Let $\cX$ be the symbol set of a constellation of order $2^m$, 
and let $N$ denotes the number of channel uses. 
In a ML-PCM scheme, 
there are $m$ component polar codes, each of length $N$.
For encoding, a length $mN$ binary vector carrying both the data bits 
and the frozen bits is split into $m$ vectors of equal length, 
and encoded by those $m$ component polar codes respectively.
Let $\bfc_j=(c_{j1},c_{j2},\cdots,c_{jN})$ denote the encoder output 
of the $j$-th component polar code for $j = 1,2,\cdots,m$.
The modulator then map the $m$-tuple 
$(c_{1i},c_{2i},\cdots,c_{mi})$ into a constellation symbol 
for transmission for $i = 1,2,\cdots,N$.
In such a way, 
each component polar code only appears at a corresponding single bit level 
for every channel use.
%
\begin{figure}[t]
    \begin{center}
    \scalebox{0.70}{%
    \begin{tikzpicture}[>=stealth,thick]
    \def\h{3.6}
    \def\w{0.7}
    \def\hh{1.0}
    \def\ww{1.5}
    \def\al{0.7}
    \def\sal{0.5}
    \def\www{0.7}
    \draw (-1.2,\h/2) node {\shortstack{Channel\\ Outputs}};
    \draw [->] (-0.5,\h/2) -- (0,\h/2);
    \draw (0,0) rectangle (\w,\h);
    \draw (\w/2,\h/2) node 
    [rotate=-270] 
    {Parallel Demodulator};
    \draw [->] (\w,\h/2) -- (\w+\sal,\h/2);
    \draw (\w+\sal,0) rectangle (\w+\sal+\www,\h);
    \draw (\w+\sal+\www/2,\h/2) node 
    [rotate=-270] 
    {Interleaver};
    \draw [->] (\w+\www+\sal,\h/2) -- (\w+\sal+\www+\sal,\h/2);
    \draw (\w+\sal+\www+\sal,\h/2-\hh/2) rectangle (\w+\sal+\www+\sal+\ww,\h/2+\hh/2);
    \draw (\w+\sal+\www+\sal+\ww/2,\h/2) node {\shortstack{Polar\\ Decoder}};
    \draw [->] (\w+\sal+\www+\sal+\ww,\h/2) --
    (\w+\sal+\www+\sal+\ww+\sal,\h/2);
    \begin{scope}[xshift = 190, yshift = -20]
    \def\h{5.0}
    \def\w{0.7}
    \def\hh{0.8}
    \def\ww{1.7}
    \def\al{0.7}
    \draw (-1.2,\h/2) node {\shortstack{Channel\\ Outputs}};
    \draw [->] (-0.5,\h/2) -- (0,\h/2);
    \draw (0,0) rectangle (\w,\h);
    \draw (\w/2,\h/2) node 
    [rotate=-270]
    {Multi-stage Demodulator};
    \foreach \i/\j in {0/1,-44/2}
    {
    \begin{scope}[yshift = \i]
        \draw [->] (\w,\h-\hh/2) -- (\w+\al,\h-\hh/2);
        \draw [->] (\w+\al+\ww,\h-\hh/2) -- (\w+\al+\ww+\al,\h-\hh/2);
        \draw [-,dashed] (\w+\al+\ww+\al/2,\h-\hh/2) -- (\w+\al+\ww+\al/2,\h-\hh/2-\hh);
        \draw [->,dashed] (\w+\al+\ww+\al/2,\h-\hh/2-\hh) -- (\w,\h-\hh/2-\hh);
        \draw (\w+\al,\h-\hh) rectangle (\w+\al+\ww,\h);
        \draw (\w+\al+\ww/2,\h-\hh/2) node {\shortstack{Polar\\ Decoder \j}};
    \end{scope}
    }
    \foreach \i/\j in {-115/m}
    {
    \begin{scope}[yshift = \i]
        \draw [->] (\w,\h-\hh/2) -- (\w+\al,\h-\hh/2);
        \draw [->] (\w+\al+\ww,\h-\hh/2) -- (\w+\al+\ww+\al,\h-\hh/2);
        \draw (\w+\al,\h-\hh) rectangle (\w+\al+\ww,\h);
        \draw (\w+\al+\ww/2,\h-\hh/2) node {\shortstack{Polar\\ Decoder \j}};
    \end{scope}
    }
    \draw (\w+\al+\ww/2,1.6) node 
    [rotate=-270] 
    {$\cdots$};
    \end{scope}
    \end{tikzpicture}
    }
    \end{center}
    \caption{BI-PCM receiver (left); ML-PCM receiver (right) for $2^m$-ary constellation.}
    \label{fig:MLPCM_receiver}
\end{figure}
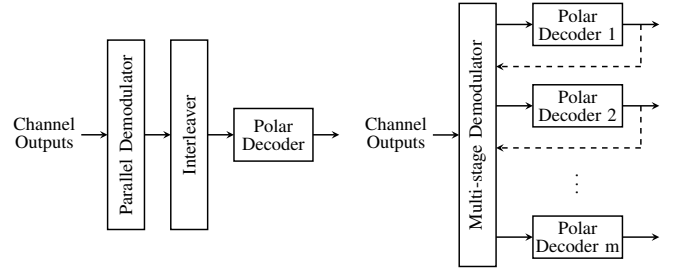

At the receiver’s side of ML-PCM, 
a multi-stage demodulator computes the soft information 
for those $m$ bit levels sequentially, 
based on both the received symbols, 
and the hard values of the previous bit levels.
More specifically, as shown in Figure \ref{fig:MLPCM_receiver} (right), 
at stage $j$ of the decoding process, 
the demodulator computes the soft information of the $j$-th 
bit level for every received symbols,
and send it to the $j$-th polar decoder. 
Then, the demodulator waits for $j$-th decoder to send back 
its decoding result. 
After retrieving the hard values for the $j$-th bit level 
of every received symbol, 
the demodulator then proceeds to the next bit level.

Let $W:\cX\rightarrow\cY$ 
be a $2^m$-ary channel with input symbol set $\cX$ with $|\cX| = 2^m$, 
and output alphabet $\cY$. 
In a ML-PCM scheme over this channel, 
$W$ is effectively decomposed 
into $m$ binary sub-channels 
preserving their mutual information.
This channel decomposition is called 
{\it sequential binary partition} (SBP) 
in \cite{seidl2013polar}, here we denote it as
$$
\psi:\; W\rightarrow\{
B_{\psi}^{(1)}, 
B_{\psi}^{(2)}, 
\cdots, 
B_{\psi}^{(m)}
\},
$$
where 
$B_{\psi}^{(j)}:\{0,1\}\rightarrow\cY\times\{0,1\}^{j-1}$ 
denotes the binary sub-channel for the $j$-th bit level 
for $j = 1,2,\cdots,m$. 
In SBP, 
each sub-channel $B_{\psi}^{(j)}$ has the 
knowledge of both the channel output $y\in\cY$, 
and their previous bit levels.
And set-partitioning (SP) labeling 
is commonly used to generate 
widely separated bit level capacities \cite{seidl2013polar}.
Let the bit-to-symbol mapping rule given by
$\cL:\{0,1\}^m\rightarrow \cX$, 
then $B_{\psi}^{(j)}$ has the transition probability
$$
B_{\psi}^{(j)}(y,b_1^{j-1}|b) = 
\frac{1}{2^{m-j}}
\sum_{b_{j}^m\in\{0,1\}^{m-j+1}:b_j=b} W(y|\cL(b_1^m))
$$
for $j = 1,\cdots,m$. 
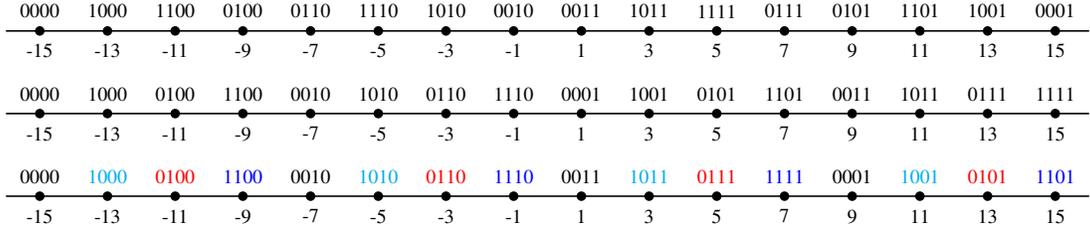
\begin{figure*}[t!]
    \begin{center}
    \scalebox{0.75}{%
    \begin{tikzpicture}[>=stealth,thick]
    \def\h{0.6}
    \def\gap{0.1}
    \draw (-16*\h,0) -- (16*\h,0);
    \foreach \i in {-15,-13,...,15}
    {
        \draw [fill] (\i*\h,0) circle (2pt);
        \draw (\i*\h,-\gap) node [below] {\i};
    }
    \draw ( 15*\h,\gap) node [above] {0001};
    \draw ( 13*\h,\gap) node [above] {1001};
    \draw ( 11*\h,\gap) node [above] {1101};
    \draw (  9*\h,\gap) node [above] {0101};
    \draw (  7*\h,\gap) node [above] {0111};
    \draw (  5*\h,\gap) node [above] {1111};
    \draw (  3*\h,\gap) node [above] {1011};
    \draw (  1*\h,\gap) node [above] {0011};
    \draw ( -1*\h,\gap) node [above] {0010};
    \draw ( -3*\h,\gap) node [above] {1010};
    \draw ( -5*\h,\gap) node [above] {1110};
    \draw ( -7*\h,\gap) node [above] {0110};
    \draw ( -9*\h,\gap) node [above] {0100};
    \draw (-11*\h,\gap) node [above] {1100};
    \draw (-13*\h,\gap) node [above] {1000};
    \draw (-15*\h,\gap) node [above] {0000};
    \end{tikzpicture}
    }
    \end{center}
    \begin{center}
    \scalebox{0.75}{%
    \begin{tikzpicture}[>=stealth,thick]
    \def\h{0.6}
    \def\gap{0.1}
    \draw (-16*\h,0) -- (16*\h,0);
    \foreach \i in {-15,-13,...,15}
    {
        \draw [fill] (\i*\h,0) circle (2pt);
        \draw (\i*\h,-\gap) node [below] {\i};
    }
    \draw ( 15*\h,\gap) node [above] {1111};
    \draw ( 13*\h,\gap) node [above] {0111};
    \draw ( 11*\h,\gap) node [above] {1011};
    \draw (  9*\h,\gap) node [above] {0011};
    \draw (  7*\h,\gap) node [above] {1101};
    \draw (  5*\h,\gap) node [above] {0101};
    \draw (  3*\h,\gap) node [above] {1001};
    \draw (  1*\h,\gap) node [above] {0001};
    \draw ( -1*\h,\gap) node [above] {1110};
    \draw ( -3*\h,\gap) node [above] {0110};
    \draw ( -5*\h,\gap) node [above] {1010};
    \draw ( -7*\h,\gap) node [above] {0010};
    \draw ( -9*\h,\gap) node [above] {1100};
    \draw (-11*\h,\gap) node [above] {0100};
    \draw (-13*\h,\gap) node [above] {1000};
    \draw (-15*\h,\gap) node [above] {0000};
    \end{tikzpicture}
    }
    \end{center}
    \begin{center}
    \scalebox{0.75}{%
    \begin{tikzpicture}[>=stealth,thick]
    \def\h{0.6}
    \def\gap{0.1}
    \draw (-16*\h,0) -- (16*\h,0);
    \foreach \i in {-15,-13,...,15}
    {
        \draw [fill] (\i*\h,0) circle (2pt);
        \draw (\i*\h,-\gap) node [below] {\i};
    }
    \draw ( 15*\h,\gap) node [above] {\color{blue}1101};
    \draw ( 13*\h,\gap) node [above] {\color{red}0101};
    \draw ( 11*\h,\gap) node [above] {\color{cyan}1001};
    \draw (  9*\h,\gap) node [above] {0001};
    \draw (  7*\h,\gap) node [above] {\color{blue}1111};
    \draw (  5*\h,\gap) node [above] {\color{red}0111};
    \draw (  3*\h,\gap) node [above] {\color{cyan}1011};
    \draw (  1*\h,\gap) node [above] {0011};
    \draw ( -1*\h,\gap) node [above] {\color{blue}1110};
    \draw ( -3*\h,\gap) node [above] {\color{red}0110};
    \draw ( -5*\h,\gap) node [above] {\color{cyan}1010};
    \draw ( -7*\h,\gap) node [above] {0010};
    \draw ( -9*\h,\gap) node [above] {\color{blue}1100};
    \draw (-11*\h,\gap) node [above] {\color{red}0100};
    \draw (-13*\h,\gap) node [above] {\color{cyan}1000};
    \draw (-15*\h,\gap) node [above] {0000};
    \end{tikzpicture}
    }
    \end{center}
    \caption{16-ASK with Gray labeling (top), SP labeling (middle), 
    and Hybrid labeling with splitting parameter  2 (bottom). 
    The first bit level lies on the left.}
    \label{fig:16PAM}
    \vspace{-0.4cm}
\end{figure*}

\section{A Hybrid Scheme for Polar Coded Modulation}
\label{sec:hybrid}
\noindent
In this section, 
we propose a hybrid polar-coded modulation scheme 
that lies between BI-PCM and ML-PCM. 
Our hybrid scheme can be viewed as a 
comprehensive framework that has BI-PCM and ML-PCM 
as its two special cases.
We begin by introducing a channel decomposition 
that we refer as {\it hybrid binary partition}. 

\subsection{Hybrid Binary Partitions}
\noindent
Let $W:\cX\rightarrow\cY$ be a discrete memoryless channel 
with input symbol set $\cX$ with $|\cX| = 2^m$, 
and output symbol set $\cY$. 
We define the {\it hybrid binary partition} (HBP) 
with splitting parameter $s$ as the channel transform
$$
\psi_s:\; 
X \rightarrow
\{
B_{\psi_s}^{(1)}, 
B_{\psi_s}^{(2)}, 
\cdots, 
B_{\psi_s}^{(m)}
\}, 
$$
where $s$ is an integer between 0 and $m$, 
and $B_{\psi_s}^{(j)}$ denotes the 
decomposed binary sub-channel 
for the $j$-th bit level for $j = 1,2,\cdots,m$. 
In this channel decomposition, 
the first $s$ sub-channels have the knowledge of 
their previous bit levels 
and the rest of the $(m-s)$ sub-channels 
only have the knowledge of the first $s$ bit levels.

Formally, for $1 \le j \le s$, we have 
$$
B_{\psi_s}^{(j)}:\;
\{0,1\}\rightarrow
\cY \times \{0,1\}^{j-1}
$$
with transition probability
$$
B_{\psi_s}^{(j)}(y,b_1^{j-1}|b) = 
\frac{1}{2^{m-j}}
\sum_{b_{j}^m\in\{0,1\}^{m-j+1}:b_j=b} W(y|\cL(b_1^m))
$$
And for $s < j \le m$, we have 
$$
B_{\psi_s}^{(j)}:\;
\{0,1\}\rightarrow
\cY \times \{0,1\}^{s}
$$
with transition probability
$$
B_{\psi_s}^{(j)}(y,b_1^{s}|b) = 
\frac{1}{2^{m-s-1}}
\sum_{b_{s+1}^m\in\{0,1\}^{m-s}:b_j=b} W(y|\cL(b_1^m)).
$$

Following this definition, 
the first $s$ sub-channels in HBP with splitting parameter $s$
will be the same as the first $s$ sub-channels 
in SBP on the same $2^m$-ary channel $W$. 

We make the remark that for a given $2^m$-ary channel $W$, 
HBP with splitting parameter $s=0$ will be the same as PBP, 
and HBP with splitting parameter $s=m$ will be the same as SBP. 
Therefore, PBP and SBP can be viewed as two special cases of HBP.

\subsection{Hybrid Labeling}
\noindent
In polar coded modulation schemes, 
PBP in BI-PCM is commonly equipped with Gray labeling, 
and SBP in ML-PCM is commonly equipped with SP labeling 
\cite{seidl2013polar}. 
Since HBP is a hybrid channel transform 
that stands between PBP and SBP, 
we propose to equip it with a {\it hybrid labeling rule} 
that stands between Gray labeling and SP labeling. 

Let $\cX$ be the symbol set for a $2^m$-ary constellation, 
we describe our hybrid labeling rule 
with splitting parameter $s$ as follows:
\begin{enumerate}
    \item 
        For every symbol $x\in\cX$, the first $s$ bit levels are 
        labeled the same as the SP labeling rule. 
    \item
        For the rest of the $(m-s)$ bit levels, 
        we first partition $\cX$ into subsets, such that  
        symbols within each subset have the same bits 
        on their first $s$ bit levels. 
        Then for each subset $\cZ\subseteq\cX$, 
        we label the rest of the $(m-s)$ bit levels 
        for symbols in $\cZ$ following the Gray labeling 
        rule for the $2^{m-s}$ sub-constellation $\cZ$.
\end{enumerate}

\begin{example}
\normalfont
We illustrate this hybrid labeling rule by taking the 16-PAM 
constellation as an example.
Figure \ref{fig:16PAM} shows examples of 
three labeling rules for 16-PAM, with Gray labeling 
at the top, SP labeling in the middle, 
and hybrid labeling with splitting parameter $s=2$ 
at the bottom.

Denote the symbol set by $\cX = \{-15,-13,\cdots,15\}$. 
In the hybrid labeling with splitting parameter $s=2$, 
the first two bit levels for every $x\in\cX$ are labeled 
the same as in SP labeling. 
Then $\cX$ can be partitioned into four subsets
according to the first two bit levels:
$$
\begin{array}{ll}
\cZ_1 = \{-15,-7,1,9\}, &\cZ_2 = \{-13,-5,3,11\},\\
\cZ_3 = \{-11,-3,5,13\}, &\cZ_4 = \{-9,-1,7,15\}.
\end{array}
$$
Those four subsets are colored differently in 
Figure \ref{fig:16PAM}.
Take $\cZ_1$ for example, 
in the hybrid labeling, 
the last two bit levels for the symbols in $\cZ_1$
are labeled following the Gray labeling rule 
viewing $\cZ_1$ as a 4-PAM constellation.
\end{example}

\subsection{Hybrid Polar Coded Modulation (Hybrid-PCM)}
\noindent
Now we describe our hybrid coded modulation scheme. 
Let $\cX$ be the symbol set of a constellation of 
order $2^m$, 
and let $N$ denotes the number of channel uses. 
In our Hybrid-PCM 
with splitting parameter $s$, 
the $2^m$-ary channel is decomposed by HBP 
with splitting parameter $s$ into $m$ binary sub-channels, 
where each of the first $s$ sub-channels 
corresponds to a component polar code of length $N$, 
and the last $(m-s)$ sub-channels correspond 
to a single component code of length $(m-s)N$. 

For encoding, a length $mN$ binary vector carrying both 
the data and the frozen bits is split into $s$ length-$N$ vectors  and one single vector of length $(m-s)N$. 
Those sub-vectors are encoded by the component   codes 
respectively.
Denote by $\bfc_j=(c_{j1},c_{j2},\cdots,c_{jN})$ 
the encoder output of the $j$-th component polar code for 
$j = 1,2,\cdots,s$, 
and denote by
$$
\bfc_{s+1}=(
\bfc_{s+1,1},\;
\bfc_{s+1,2},\;
\cdots,\;
\bfc_{s+1,N}
)
$$
the encoder output of the $(s+1)$-th component   code, 
where $\bfc_{s+1,i}$ is a length-$(m-s)$ vector for $i=1,2,\cdots,N$. 
The modulator then map the $m$-tuple 
$$
(
c_{1i},\;
c_{2i},\;
\cdots,\;
c_{si},\;
\bfc_{s+1,i})
$$
into a constellation symbol 
for transmission for $i = 1,2,\cdots,N$.

\begin{figure}[t]
    \begin{center}
    \scalebox{0.7}{%
    \begin{tikzpicture}[>=stealth,thick]
    \def\h{8}
    \def\w{1}
    \def\hh{0.55}
    \def\ww{2.6}
    \def\al{1.1}
    \draw (-1.2,\h/2) node {\shortstack{Channel\\ Outputs}};
    \draw [->] (-0.5,\h/2) -- (0,\h/2);
    \draw (0,0) rectangle (\w,\h);
    \draw (\w/2,\h/2) node 
    [rotate=-270]
    {Multi-stage Demodulator};
    \foreach \i/\j in {0/1,-33/2,-85/s}
    {
    \begin{scope}[yshift = \i]
        \draw [->] (\w,\h-\hh/2) -- (\w+\al,\h-\hh/2);
        \draw [->] (\w+\al+\ww,\h-\hh/2) -- (\w+\al+\ww+\al,\h-\hh/2);
        \draw [-,dashed] (\w+\al+\ww+\al/2,\h-\hh/2) -- (\w+\al+\ww+\al/2,\h-\hh/2-\hh);
        \draw [->,dashed] (\w+\al+\ww+\al/2,\h-\hh/2-\hh) -- (\w,\h-\hh/2-\hh);
        \draw (\w+\al,\h-\hh) rectangle (\w+\al+\ww,\h);
        \draw (\w+\al+\ww/2,\h-\hh/2) node {Polar Decoder \j};
    \end{scope}
    }
    \draw (\w+\al+\ww/2,5.5) node 
    [rotate=-270] 
    {$\cdots$};
    \def\h{3.6}
    \def\w{1}
    \def\hh{1.0}
    \def\ww{3.0}
    \def\al{0.7}
    \def\sal{0.4}
    \def\www{0.7}
    \draw [->] (\w,\h/2) -- (\w+\sal,\h/2);
    \draw (\w+\sal,0) rectangle (\w+\sal+\www,\h);
    \draw (\w+\sal+\www/2,\h/2) node 
    [rotate=-270] 
    {Interleaver};
    \draw [->] (\w+\www+\sal,\h/2) -- (\w+\sal+\www+\sal,\h/2);
    \draw (\w+\sal+\www+\sal,\h/2-\hh/2) rectangle (\w+\sal+\www+\sal+\ww,\h/2+\hh/2);
    \draw (\w+\sal+\www+\sal+\ww/2,\h/2) node {Decoder s+1};
    \draw [->] (\w+\sal+\www+\sal+\ww,\h/2) --
    (\w+\sal+\www+\sal+\ww+\sal,\h/2);
    \end{tikzpicture}
    }
    \end{center}
    \caption{Hybrid-PCM receiver 
    with splitting parameter  $s$}
    \label{fig:HPCM_receiver}
    \vspace{-0.5cm}
\end{figure}
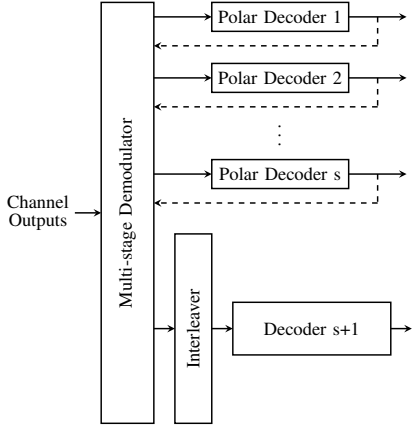
At the receiver's sided of Hybrid-PCM, 
as shown in Figure \ref{fig:HPCM_receiver},
a multi-stage decoder first computes the 
soft information for the first $s$ bit levels sequentially, 
based on both the received symbols, 
and the hard values of the previous bit levels.
Then the demodulator computes the soft information 
for the rest of the $(m-s)$ bit levels in parallel, 
based on the received symbols, 
and the hard values of the first $s$ bit levels.
The soft information for the last $(m-s)$ bit 
levels is then de-interleaved, and then fed 
to the decoder for the last component code.
In our hybrid scheme, we employ the hybrid labeling rule 
with the same splitting paramete $s$ for 
the mapping between the coded bits and the constellation symbols.

Note that for a $2^m$-ary channel $W$, 
this hybrid scheme with the splitting parameter $s$ can be viewed as a general framework that includes both BI-PCM and ML-PCM as special cases by setting $s=0$ and  $s=m$, respectively. 
\section{Performance Evaluation}\label{sec:sim}
\noindent
We present some simulation results of our 
hybrid scheme on 64-QAM and 256-QAM where all polar codes  
are constructed following the Monte Carlo construction.

\begin{figure}[t]
\centering
\includegraphics[trim=60 210 60 210, width=3.15in]{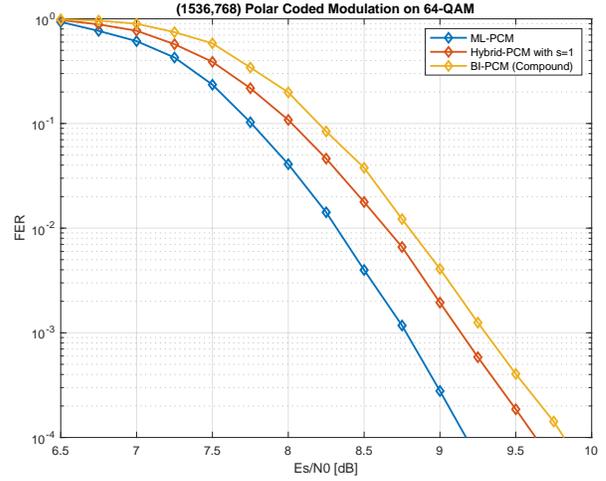}
\caption{Performance comparison for 
ML-PCM, Hybrid-PCM and BI-PCM with compound polar code on 64-QAM.}
\label{fig:64QAM}
\end{figure}

Figure \ref{fig:64QAM} shows the simulation results 
of SC decoding for $(1536,768)$ polar coded modulation 
on 64-QAM with three difference schemes: 
ML-PCM, Hybrid-PCM with splitting parameter $s=1$, 
and BI-PCM with compound polar code. 
In our experiment, 
every 64-QAM is simulated by two independent 8-PAM symbols, 
and all the coded modulation schemes 
are applied on the 8-PAM constellation.
We can observe that ML-PCM performs better than Hybrid-PCM, 
and BI-PCM with compound polar code as expected, 
and our hybrid scheme with splitting parameter $s=1$ 
shows an approximated 0.5 dB performance gain over BI-PCM with compound polar code. 

\begin{figure}[t]
\centering
\includegraphics[trim=60 210 60 210, width=3.15in]{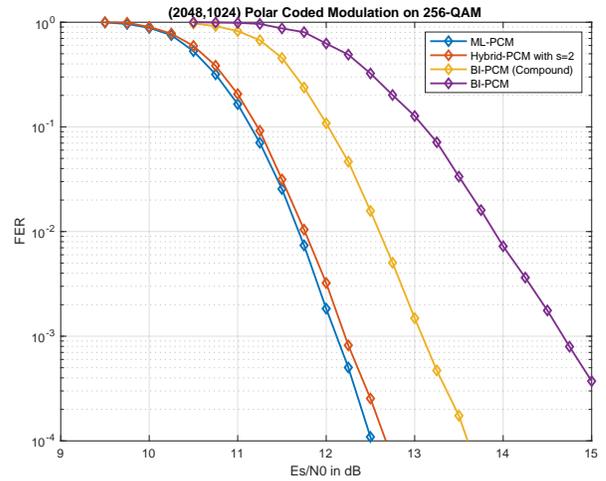}
\caption{Performance comparison for ML-PCM, Hybrid-PCM, 
BI-PCM with compound polar code, and plain BI-PCM 
on 256-QAM}
\label{fig:256QAM}
\end{figure}

Figure \ref{fig:256QAM} shows the simulation results 
of SC decoding for $(2048,1024)$ polar coded modulation 
on 256-QAM with four difference schemes: 
ML-PCM, Hybrid-PCM with splitting parameter $s=2$, 
BI-PCM with compound polar code, 
and plain BI-PCM (Figure \ref{fig:MLPCM_receiver}). 
In our experiment, 
every 256-QAM is simulated by two independent 16-PAM symbols, 
and all the coded modulation schemes 
are applied on the 16-PAM constellation.
We can see that our hybrid scheme 
with splitting parameter $s=2$ can close up 
majority of the 
performance gain of ML-PCM over BI-PCM with compound polar code, 
while reducing the number of required iterative information exchanges 
between the demodulator and the decoder in ML-PCM by half, thus reducing the overall decoding latency.

\section{Conclusion and Discussion}
\label{sec:conc}
\noindent
We have proposed a new  polar coded modulation scheme that uses hybrid bit partitions by assigning only a fraction of bits for {\it sequential binary partition} and subsequent iterative demodulation and decoding, whereas the remaining bits are subject to {\it parallel binary partition} and corresponding parallel demodulation and then decoding. It can be viewed as a comprehensive framework that includes both ML-PCM and BI-PCM as special cases, and our simulation results have shown that it can alleviate the latency in ML-PCM 
while still maintaining a considerable performance gain 
over BI-PCM.

Although we only discussed polar codes as 
component codes for our hybrid coded modulation, 
in principle, just like MLC and BICM, 
any other codes such as Turbo or LDPC codes 
can also be chosen as component codes in our hybrid scheme.
The flexibility of working together with other codes, reduced latency compared to ML-PCM, and the large performance gain over BI-PCM on high order modulation
make the proposed hybrid polar coded modulation scheme attractive for future
communication systems such as 6G.




\newpage
\bibliographystyle{IEEEtran}
\bibliography{refs}

\end{document}